# Multi-level Governance, Smart Meter Adoption, and Utilities' Energy Efficiency Savings in the U.S.

Yue Gao*,
Assistant Teaching Professor,
 School of Business,
*Clark University, USA*
yugao@clarku.edu

Jing Zhang
Full Professor,
School of Business,
*Clark University, USA*
jizhang@clarku.edu

ABSTRACT

Smart grids enable the alignment of energy supply and demand, enhance energy efficiency, and raise consumer awareness of energy conservation. Smart meter, a vital technological component of smart grids, enables bidirectional communication between consumers and utility companies. This paper employs the two-stage least squares panel data model to examine the effects of federal funding and state legislative actions on smart meter adoption and the resulting energy efficiency savings in the residential sector of the United States. Additionally, we use machine learning techniques to select the subset of control variables that are influential for smart meter adoption and energy efficiency savings. The findings suggest that both federal financial assistance and state legislative actions have positive effects on smart meter adoption, and the interaction of federal funding and state policy interventions have a significantly greater effect on adoption as compared to the sum of the individual policy instrument alone. Furthermore, there exists a positive association between the adoption rate and energy efficiency savings. This study presents empirical evidence that underscores the significance of multi-level governance as an effective means for policy integration regarding smart meter adoptions, and it discusses the policy implications for grid modernization and environmental policies.

*Corresponding author

# 1 Introduction

As a critical infrastructure component, smart grids facilitate the development of smart cities and smart governance. The alignment of energy supply and demand is a fundamental aspect of this technological advancement (Gil-Garcia, Zhang, & Puron-Cid, 2016; Mah, van der Vleuten, Ip, & Hills, 2012). Consequently, the adoption of smart grids has gained significant attention in the field of digital government research (Gungor et al., 2011; D. Zhang, Pee, Pan, & Cui, 2022). The U.S. Department of Energy (DOE) defines the smart grid as an intelligent electricity grid that enables two-way communication between utility companies and their customers, as well as the ability to sense activity along transmission lines (DOE, 2021). By integrating information and communication technologies (ICTs) with the existing electricity infrastructure, the smart grid enhances the cost-effectiveness and resilience of the electricity network. It is widely recognized as a crucial enabler for achieving energy independence and promoting low-carbon economic growth (Strong, 2019). As the share of renewable energy sources like solar and geothermal power continues to rise, the intermittent nature of their generation poses challenges for the traditional power grids in accommodating them. Smart grids, however, are specifically designed to integrate diverse power generation sources seamlessly. Furthermore, smart grids offer the potential for more efficient energy consumption by reducing power outages, minimizing delivery costs, and encouraging energy-conscious behaviors among consumers (ESGTF, 2010). One key technological component of a smart grid is the smart meter, also referred to as advanced metering infrastructure (AMI). It facilitates real-time or near real-time communication between consumers and energy companies, enabling the exchange of usage data (EPRI, 2007; FERC, 2019 ).

The deployment of smart meter technologies in conjunction with customer-based systems enables utilities to offer demand-side management programs that motivate customers to reduce their electricity consumption without significantly sacrificing the services provided (DOE, 2016;

Lang & Okwelum, 2015). One example of such programs offered by municipal utilities in Massachusetts is customers may claim a rebate of up to $15,000 for installing high-efficiency heat pumps (MassSave, 2022).

The implementation of the smart grid in the United States encounters several challenges, despite its numerous advantages. One significant issue resolves around the scarcity of financial support (Guo, Li, & Lam, 2017). It is widely recognized that this undertaking demands substantial capital investment and entails a lengthy investment cycle (Depuru, Wang, Devabhaktuni, & Gudi, 2011). To establish a fully functional smart grid in the country, an estimated net investment ranging from $338 to $476 billion is projected (Gellings, 2011). Feng, Zhang, and Gao (2016) argue that certain smart grid projects may not be economically viable without adequate financial assistance, which can be attained through a combination of public funds and private investments. Other challenges arise from the absence of technology standards and regulatory frameworks, which can potentially hinder private investment and create risks and uncertainties (DOE, 2020). Compared to traditional meters, the adoption of AMI generates a substantial volume of consumer data. This development gives rise to new challenges pertaining to the security and accessibility of customer data, as well as the protection of data privacy from cybersecurity breaches and other threats (ACEEE, 2020). Moreover, in a modern grid, ensuring interoperability is essential, as it enables safe and efficient sharing and utilization of information among multiple networks, systems, devices, applications, or components with minimal constraints (DOE, 2016).

To overcome these challenges, public policies often play a critical role in accelerating such technology deployment (Giest, 2020; Horbach, Rammer, & Rennings, 2012; Norberg-Bohm, 2000; Ockwell, Watson, MacKerron, Pal, & Yamin, 2008; F. Zhang & Gallagher, 2016). The potential of smart grids has led governments worldwide to acknowledge their importance and consider them as a crucial investment. Consequently, significant public financing and legislative initiatives have been directed toward the adoption of smart grids (Mah et al., 2012). An illustrative example is the European Union (EU), where member states had planned to allocate a large portion, specifically 84%, of the $1.147 billion investment in energy infrastructure projects to electricity and smart grid initiatives by 2020 (Power Technology, 2020). Similarly, the China Electricity Council reported an accumulated investment of around 0.63 trillion USD for the smart grid in China by 2020 (Globe News Wire, 2020).

In the U.S., the policy environment governing the electricity sector is characterized by a high degree of complexity, involving multiple public institutions that possess independent but interconnected regulatory jurisdictions. At the federal level, DOE establishes general policies, while the Environmental Protection Agency (EPA) focuses on environmental regulations, and the Federal Trade Commission (FTC) is responsible for developing and enforcing consumer protection policies (EPR, 2008). In terms of economic regulation within the electric power distribution segment, individual states typically rely on their Public Utilities Commissions (PUC) to oversea this aspect, whereas the Federal Energy Regulatory Commission (FERC) governs the interstate transmission segment (EPR, 2008; FERC, 2022). Furthermore, some states approve pilot projects and full-scale deployments of smart meters, taking into consideration the most appropriate cost recovery mechanisms (FERC, 2010). A practical illustration of this can be seen in the New York Public Service Commission's approval of a surcharge for Consolidated Edison Company's smart grid demonstration project as well as deferred cost recovery for five other utility projects funded by ARRA, with the intention of reviewing them in future rate cases (New York Public Service Commission, 2010). Given the increasing concerns surrounding consumer privacy associated with AMI, states have also taken steps to address this issue. In 2010, for instance, California enacted laws aimed at safeguarding consumers' privacy and protecting their energy consumption data (California Legislative Counsel, 2009-2010).

Therefore, to better understand the connections between government interventions and grid modernization, it is important to incorporate a multi-level governance approach (Bakker 2018), especially with regard to the dynamic interactions between federal and state policies. Previous research in this area has been limited, with few studies adopting a polycentric governance lens to explore the driving forces behind advanced metering adoption. Moreover, these studies have not incorporated the most recent data and have focused primarily on state-level analyses rather than examining utilities at a more granular level. (e.g., Gao, Fang, & Zhang, 2022; Zhou & Matisoff, 2016). Therefore, a comprehensive assessment of the impacts on the grid operations, such as energy efficiency, which represents a higher-level objective for policy interventions and smart meter adoption, has been lacking.

This study attempts to examine the effectiveness of multi-level governance in promoting the adoption of smart grid technology, and the subsequent energy efficiency savings achieved by utilities. To investigate this, we employ longitudinal data and utilize machine learning and econometrics models. Our research findings demonstrate that federal funds, state legislative activities, as well as their interaction, have a positive connection with the smart meter adoption, which in turn, leads to greater energy efficiency savings in the U.S. residential sector. These results provide a more comprehensive understanding of the interplay between multi-level governance, smart meter adoption, and energy efficiency. The study contributes to both theoretical advancements in smart meter technology adoption and practical implications for public investment and regulatory frameworks.

# 2 Literature review

## 2.1 Smart meter technologies and adoption

Until 2013, automatic meter reading (AMR), which involved one-way communication from meters to utilities, was the predominant method for collecting consumption data. This data was then transmitted either to a remote collector in a utility employee's vehicle or to a central location through a fixed network (EIA, 2017; Strong, 2019). However, with the advancements in digital technology, the concept of smart meters, or AMI, has emerged, transforming communication and data systems from one-way to two-way (An, 2011). According to the U.S. Energy Information Administration (EIA), as of 2020, there were approximately 102.9 million AMIs installed in electric utilities across the United States, with 88% of them being deployed in the residential sector. Alongside the previously mentioned benefits, such as enabling demand response and facilitating distributed generation, AMI systems offer utilities precise information during storms or other system failures. This accurate outage data empowers utilities to restore service more efficiently and reduce the average duration of electric system outages.

Figure 1 illustrates the overall trend of smart meter installation in the U.S. residential sector between 2007 and 2020. Notably, in 2013, the number of AMI meters surpassed that of AMR meters in the residential sector for the first time. By the end of 2020, 90 million meters were operating in AMI mode, while 29 million meters remained in AMR mode (EIA, 2007-2019). Furthermore, Figure 2 presents the spatial distribution of AMI adoption rates at the state level in the U.S. residential sector in 2020. The degree of residential AMI adoption varies significantly among states. Among all states, three states boasted an AMI penetration rate exceeding 95 percent. The District of Columbia achieved the highest AMI penetration rate at 99.8 percent, followed by Pennsylvania at 98.3 percent and Nevada at 96.5 percent. Additionally, six other states reported residential AMI adoption rates surpassing 90 percent in 2020, namely Michigan, Kansas, Maine, Illinois, Tennessee, and Georgia (EIA, 2007-2019).

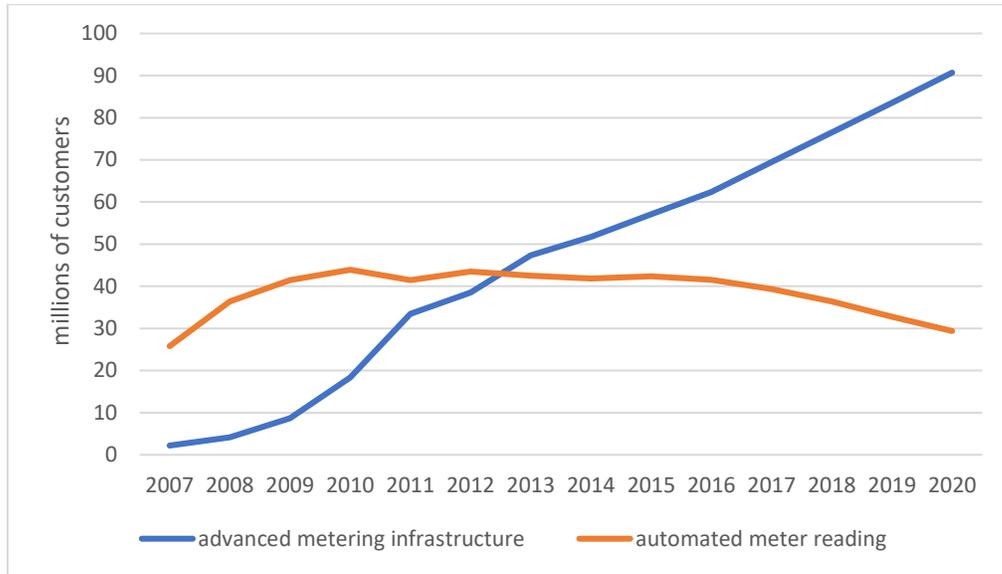

**Figure 1.** U.S. advanced meter and automatic meter adoption in the residential sector, 2007-2020. (Note: this figure is based on data from U.S. EIA)

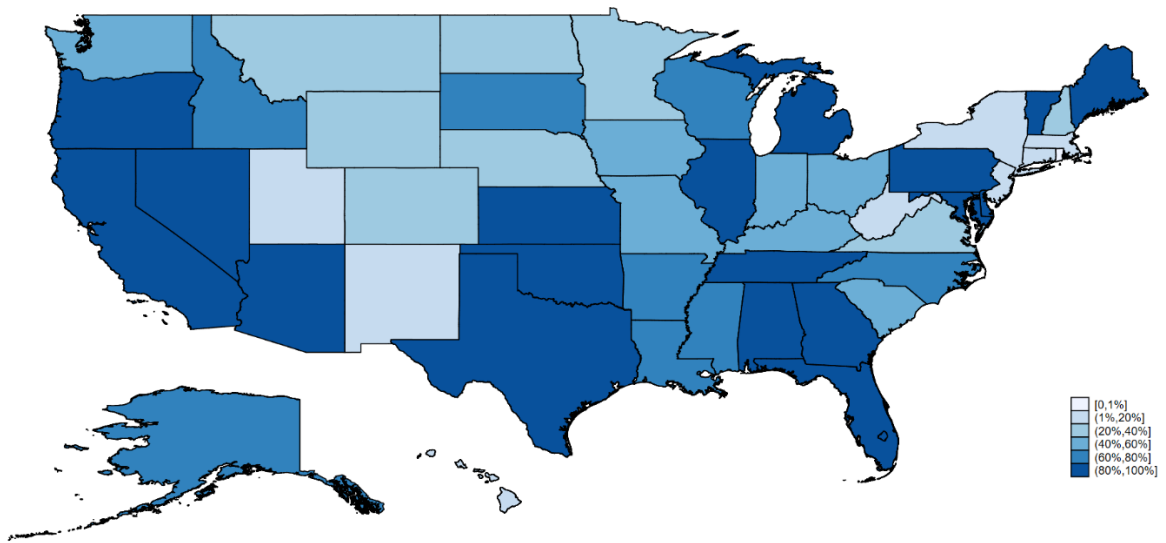

**Figure 2.** U.S. residential AMI adoption rates by state, 2020 (Note: this figure is based on data from U.S. EIA)

According to EIA, the residential sector has consistently accounted for the highest percentage of total revenue in the electricity industry from 2007 to 2020. The commercial and industrial sectors follow as the second and third largest contributors, respectively. In 2020, residential customers, numbering 136.68 million, directly consumed 1,464.6 Terawatt hours, reporting 39.4% of the total consumption valued at 192 billion (EIA, 2020). This research focuses specifically on the residential sector due to its significant market share in power consumption. The decisions made by households collectively have a great impact on energy efficiency, making it essential to understand the effectiveness of policy instruments on energy consumption. In addition, it is important to recognize that the motivations and decision-making processes of residential and commercial actors can differ significantly. For example, residential electricity demand tends to peak during hot summer afternoons due to increased air conditioning use, while commercial sector demand is highest during business operating hours and decreases significantly during nights and weekends (EPA, 2014). Therefore, it is crucial to treat these sectors differently to gain insights into their unique characteristics and tailor policy interventions accordingly.

2.2 Policy factors of smart meter adoption

To accelerate the adoption of clean energy technology adoption, government interventions have emerged as crucial factors in addressing various obstacles (Jaffe, Newell, & Stavins, 2005; Jaffe & Stavins, 1994). These obstacles encompass challenges such as inadequate funding, absence of technological standards, heterogeneity among potential adopters, information deficiency, and regulatory uncertainty (Jaffe & Stavins, 1994). Extensive research has explored a range of policy instruments that play pivotal roles in facilitating clean energy adoption. These instruments include financial incentives in the form of grants or tax benefits, establishment of regulatory goals, implementation of technology standards, and initiatives to address information gaps through demonstration projects and advertising campaigns (Howlett, Vince, & Del Rio, 2017; Jaffe et al., 2005; Norberg-Bohm, 2000).

The federal government has implemented various measures to promote the deployment of smart meters in the United States. One such measure is the Energy Policy Act of 2005 (EPACT), which advises utility regulators to explore demand response programs and mandates utilities to provide customers with time-based rate schedules and meters (U.S. Congress, 2005). Time-based rate schedules enable customers to manage their energy consumption and costs using advanced metering and communication devices (U.S. Congress, 2005). While it is not compulsory, each state regulatory authority is required to assess the feasibility of implementing such policies within its jurisdiction. Additionally, EPACT stipulates that the DOE must inform consumers about the availability and benefits of advanced metering and communication technologies, and provide financial support for demonstration or pilot projects. The Energy Independence and Security Act of 2007 (EISA) further directs DOE, the Federal Energy Regulatory Commission (FERC), states, and utilities to establish programs that encourage the installation of smart metering (U.S. Congress, 2007). Following the enactment of these acts, a survey conducted in 2008 on state regulatory actions revealed that 38 state utility commissions had initiated regulatory discussions regarding smart meters and demand response in line with federal legislation, with 32 states having completed the process (Faruqui, Hledik, & Sergici, 2009). Under EISA, the National Institute of Standards and Technology (NIST) and FERC are responsible for developing and deploying smart grid technological and interoperability standards. In 2010, NIST released the NIST Framework and Roadmap for Smart Grid Interoperability Standards, Release 1.0, which highlighted the need for numerous standards, specifications, and requirements for smart grids (FERC, 2010). In line with its mandate, NIST launched the Smart Grid Interoperability Panel (SGIP) 2.0 in 2013 to support the development of standards for smart grids (FERC, 2019; NIST, 2021).

The American Recovery and Reinvestment Act of 2009 (Recovery Act) stands out as a notable intervention in 2009 in the promotion of smart meter adoption. This act allocated 4.5 billion to the DOE to expedite the modernization of the electric power grid. As a result, the Smart Grid Investment Grant (SGIG) program, DOE, and the electricity industry have collectively invested $8 billion in 99 cost-shared projects involving over 200 participating electric utilities. Participating electric utilities were selected through merit-based, competitive solicitations that emphasized factors such as technology, project plan, methods for addressing interoperability and cybersecurity, etc. The SGIG program invests in setting up existing smart grid technologies, tools, and techniques to improve grid performance. The effect of this program has been documented as 'increased flexibility, functionality, interoperability, cybersecurity, situational awareness, and operational efficiency' (DOE, 2021). Several empirical studies have examined the impacts of the policy factors on the adoption of smart meters. Zhou and Matisoff (2016) and Gao et al. (2022) have found compelling evidence supporting the role of federal financial incentives as a primary driver of smart meter adoption at the state level. In light of these findings, we propose Hypothesis 1

**Hypothesis 1**: The presence of federal government funding leads to a higher rate of smart meter adoption by utilities.

Second, variations in the diffusion rate of smart meters are also often driven by state legislations and regulations (Gungor et al., 2011, Sovocool et al. 2021). State governments, through legislative branches and Public Utilities Commissions (PUCs), take actions to support or mandate the deployment of smart meters. These actions are necessary as smart meter deployment typically involves the approval of utility infrastructure investments and the establishment of time-variant power tariffs or regulatory rulings. State-level legislative actions encompass various measures, including, but not limited to, 1) requirements for cost-benefit analysis, where some PUCs mandate a positive cost-benefit analysis of advanced metering investment (DOE, 2020; FERC, 2019), while others provide flexibility in creating a positive business case based on intangible benefits; 2) provision of 'opt-out' options, allowing customers in certain states to choose not to have advanced meters installed in their homes (Shea & Bell, 2019); 3) adoption of time-variant pricing, as observed in states such as California and Arizona with the implementation of Time-of-Use pricing; and 4) the establishment of data privacy and security policies, approved by some PUCs (DOE, 2020; EIA, 2011).

    Previous studies have shed light on the importance of state policies in driving smart meter adoptions. For instance, Zhou and Matisoff (2016) found that state legislative and regulatory activities play a crucial role in increasing the penetration rate of smart meters. Strong (2019), employing discrete duration models and fractional response models, found that policy and regulatory support contribute to higher levels of smart meter penetration. Similarly, Y. Gao et al. (2022), through spatial analysis, revealed a positive impact of state legislative actions on smart meter adoption at the state level. However, these studies also highlighted certain limitations and suggested avenues for future research. Strong's analyses failed to consider the influence of federal funding or its interaction with state regulatory support and did not quantitatively assess the extent of impacts, relying solely on a dummy variable to measure state policy. S. Zhou and Matisoff (2016) utilized a dataset covering the period from 2007 to 2012, omitting after the post-2013 period when the effects of the ARRA Act came into play. Their study, along with Y. Gao et al. (2022), employed aggregated data at the state level, thus failing to capture within-state variation and examine the impacts of smart meter adoption on utility performance. Furthermore, Zhou and

Matisoff's (2016) approach to measure state policy strength with a simple dummy variable is suboptimal.

Recent research conducted by Matisoff(2008), Schaffrin, Sewerin, and Seubert(2015), Viscusi and Hamilton(1999), and Zhou and Matisoff(2016) has demonstrated a positive correlation between policy count and policy strength Consequently, for measurement of state legislative activities pertaining to advanced metering, we adopt a policy counts approach. Specifically, our analysis focuses on state legislation, as well as orders and requirements issued by PUC. We exclude government reports, proposals, recommendations, or policy analyses, as their impact may not be uniform across all utilities. To ensure comprehensive coverage, we sourced state policy data from various reputable sources, including DOE(2020), FERC(2019), S. Zhou and Matisoff(2016). In light of these considerations, we propose our second hypothesis as follows,

**Hypothesis 2**: A greater number of state legislative actions lead to a higher level of smart meter adoption by utilities.

The concept of 'multilevel governance' (MLG) was initially developed by Liesbet Hooghe and Gary Marks in the early 1990s and has since been extensively studied in the context of European integration. MLG refers to a system of 'continuous negotiation among nested governments at several territorial tiers – supranational, national, regional and local" (Marks, 1993), all entangled in territorially encompassing policy networks (Bache 2005). In the context of European integration, MLG theory explores the increasing frequency and complex interactions among government actors, as well as the growing importance of provincial, state, and municipal governments in shaping cohesion policies, particularly in the integration of climate actions (Bache, 2005). In practice, MLG within the EU emphasizes coordinated actions between all levels of government in policymaking, fulfillment, and assessment of European policies (Coopenergy Consortium, 2015).

Similarly, the complexity of the multi-layer governance structure in the U.S. requires scholars to go beyond the independent influences of federal and state policies, and consider the possible interaction between multiple government units (Andersson & Ostrom, 2008; Feldman, 2016; E. Ostrom, 2009; V. Ostrom, Tiebout, & Warren, 1961). According to Shobe and Burtraw (2012), the effectiveness of policies established by multiple tiers of government with responsibility for the same area might enhance or attenuate the overall effectiveness of the policies depending on how successful these policies interact. The issues and challenges of designing a mix of policy tools in sustainable development have been well documented in a number of areas, such as demand-side management (Howlett et al., 2017; J. Zhang, Luna-Reyes, Pardo, & Sayogo, 2016). In the U.S., literature on the impacts of a mix of policy tools has revealed both positive as well as negative interactions, depending on the specific context. In the example of climate change policy (Goulder & Stavins, 2011), the overlap of state and federal regulations results in problematic interactions in the context of greenhouse gas reduction. State actions may fail to lower greenhouse gas emissions nationally and reduce the cost-effectiveness of the total national effort. Using a price-based (rather than a quantity-based) federal policy could have avoided the problems caused by overlapping regulations (del Río & Cerdá, 2017; Goulder & Stavins, 2011). Alternatively, they may interact positively, such as with affordable health insurance and highway systems. Under the Affordable Care Act (ACA), the federal government offers the majority of the funding for subsidized coverage and establishes a floor for healthcare insurance market regulation, states have the flexibility to manage their Medicaid program and run their insurance marketplace. The ACA's combination of federal standards and subsidies, along with state regulatory power, dramatically escalates coverage and access across the nation, while also narrowing regional disparities (The Commonwealth Fund, 2020). In the context of smart meter adoption, studies have shown that a polycentric government system, characterized by interactions among federal financial incentives, state policies, and regulatory uncertainty, drives technological change in advanced metering in the U.S. (S. Zhou and Matisoff , 2016). Furthermore, a comparative case study examining smart meter policy schemes in five European countries identified the presence of policy bundles addressing multiple barriers to smart meter adoption as a differentiating factor between leaders and laggards in diffusion (S. Zhou and Brown, 2017). Thus, we propose the following hypothesis:

**Hypothesis 3**: The interaction between the presence of federal government funding and state legislative actions leads to a higher level of smart meter adoption by utilities.

## 2.3 Energy Efficiency Effects of Smart Meter Adoption

One of the key advantages of smart grid adoption for consumers and utilities is the potential for increased energy efficiency (DOE, 2010; L. Zhou, Xu, & Ma, 2010). Energy efficiency, as defined by the EIA, refers to the reduction of energy used by end-use devices and systems while maintaining the same level of services provided (EIA, 2007-2019). By connecting AMI to various information and management systems, utilities can unlock a range of new features that enhance the efficiency of grid operations (DOE, 2016). Firstly, the implementation of AMI enables improved overall system operation efficiency, resulting in reduced electricity rates. This is achieved through the adoption of more efficient procedures that surpass existing standards for energy efficiency measures. Secondly, by providing consumers with access to enhanced information, they can make more informed decisions regarding their electricity usage (GAO, 2011; C. Guo, Bond, & Narayanan, 2015; Moura, López, Moreno, & De Almeida, 2013). Consequently consumers can make energy efficiency purchase decisions and modify their usage behaviors accordingly (Corbett, Wardle, & Chen, 2018; Olmos, Ruester, Liong, & Glachant, 2011; L. Zhou et al., 2010). For example, consumers can opt to replace light bulbs or outdated HVAC systems with high-efficiency devices, contributing to energy savings (EIA, 2007-2019).

Several previous empirical studies have investigated the impacts of smart meter adoption on the energy performance of utility companies. However, these studies have certain limitations that need to be addressed. For instance, Corbett (2018) found a positive influence of AMI meters on energy efficiency in the U.S. between 2009 and 2012. However, the analysis suffers from small sample bias and the issue of multicollinearity. In contrast, evidence from the Mediterranean region suggests that the use of smart meters effectively enhances energy efficiency in low-income households (Podgornik, Sucic, Bevk, & Stanicic, 2013). Another study by Samuels and Booysen (2019), which conducted a controlled behavioral experiment in South African schools, reveals financial savings of 11% and 14% in two treated schools that received highly visualized energy usage reports. This suggests that providing high-frequency and easily understandable information can encourage schools to reduce energy expenses (Samuels & Booysen, 2019). Additionally, certain demand response activities related to smart meters in the Austrian residential electricity sector, such as critical peak prices and simple time-of-use tariffs, have demonstrated the highest energy efficiency (Olmos et al., 2011). Similarly, Arif et al.(2013) found that overall energy consumption decreased when consumers in Saudi Arabia received notifications about their energy usage through SMS messages or the Internet, facilitated by smart meters. Although existing studies indicate a general increase in energy efficiency due to smart meter adoption, none of them have conducted a comprehensive empirical analysis or addressed issues of endogeneity and multicollinearity. Therefore, our fourth hypothesis is as follows,

**Hypothesis 4**: An increase in smart meter adoption is associated with an increase in utilities' energy efficiency savings.

In summary, this research aims to examine the effects of multi-level government policies on smart meter adoption and explore the subsequent impact of adoption on utilities' energy efficiency savings. The study focuses on the U.S. residential sector and utilizes evidence to analyze these relationships. Figure 3 presents a visual representation of the research framework and the key components being studied.

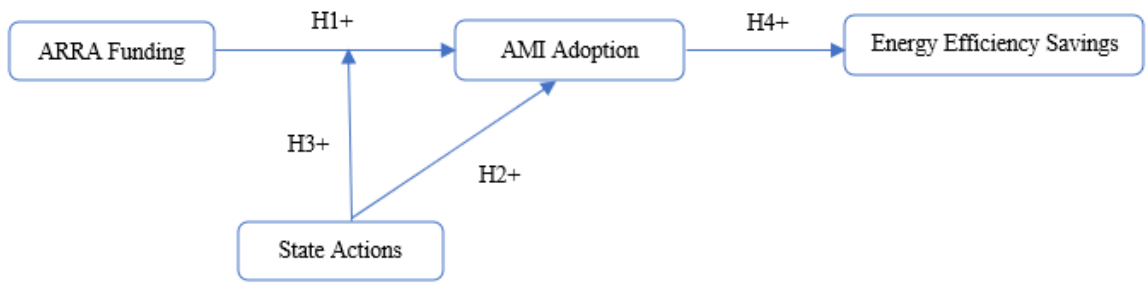

**Figure 3.** Research model on multi-level governance, smart meter adoption, and utilities' energy efficiency savings

## 3 Methodology

In our study, we employ econometric models to analyze the relationship between multi-level governance, smart meter adoption, and utilities' energy efficiency savings. However, two econometric challenges need to be addressed. Firstly, the true data-generating process is unknown, and including a large number of control variables can lead to overfitting, while using too few or the wrong ones may lead to omitted variable bias. To tackle this issue, we adopt the LASSO estimator and cross-validation approach, following the methodology proposed by Belloni, Chernozhukov, and Hansen (2014), and Drukker & Liu(2019). The LASSO regression is a machine learning technique that addresses overfitting and selects a subset of covariates by minimizing a modified cost function. It estimates the coefficients by finding the values that minimize the following equation:

$$\sum_{i=1}^{n}\left(y_i - \beta_0 - \sum_{j=1}^{p}\beta_j x_{ij}\right)^2 + \lambda \sum_{j=1}^{p}|\beta_j|$$

Here, λ is a tuning parameter that controls the level of regularization. I λ forces the coefficients of unimportant variables to be exactly zero By doing so, LASSO performs variable selection and provides parsimonious models with a subset of relevant regressors. However, selecting an appropriate value for λ is crucial. To determine the optimal value of λ, we employ cross-validation, another machine learning n. Specifically, the dataset is randomly divided into 10 non-overlapping groups of equal size. One group is designated as the validation set, while the remaining nine groups are used to fit the LASSO regression. The mean squared error (MSE) is computed on the validation set. This process is repeated nine times, with each group serving as the validation set once. The cross-validation MSE is then calculated as the average of the MSEs from the 10 groups. The ideal value of λ is chosen as the one that minimizes the cross-validation MSE (James, Witten, Hastie, & Tibshirani, 2013). By employing the LASSO estimator and cross-validation, we address the challenges of overfitting and variable selection, ensuring a robust and accurate analysis of the relationship between multi-level governance, smart meter adoption, and utilities' energy efficiency savings in our study.

    Another econometric concern pertains to the presence of confounding variables that may impact utilities' performance. Factors such as customer environmental consciousness, technology acceptance, economic sensitivity, and the quality of management and leadership within utility companies are among the unobserved variables that could be correlated with smart meter adoption. Furthermore, it is plausible that utilities' energy savings exert a causal influence on smart meter adoption. Thus, bias arises due to endogeneity resulting from omitted variables and reverse causality in the regression analysis. To address the issue of endogeneity in smart meter adoption,

we use government policies as instrument variables and utilize a two-stage least square (2SLS) method. In the first stage, given the continuous and censored nature of the dependent variable, we estimate a random effects Tobit model to regress smart meter adoption on multilevel government policies, including federal funding, state policies, and their interaction. Next, we obtain the fitted values of smart meter adoption. In the second stage, we regress energy efficiency measures on the fitted adoption. The 2SLS approach offers a significant advantage over ordinary least square models by leveraging the variation in smart meter adoption that aligns with the variation in the multi-level government policy measures. By utilizing this specific variation, the 2SLS method computes the slope estimate (Kennedy, 2008). In line with prior empirical research on related topics (Corbett et al., 2018; Strong, 2019; S. Zhou & Matisoff, 2016), statistical model specifications are identified as follows,

$AMI\_Adoption_{ijt} = \alpha_0 + \alpha_1 ARRA\_funding_{ijt} + \alpha_2 State\_Actions_{jt} + \alpha_3 ARRA\_funding_{ijt} \times State\_Actions_{jt} + \alpha_4 C_{1,ijt} + \alpha_5 C_{2,jt} + \mu_{ij} + \epsilon_{ijt}$

(1)

$Energy\_Efficiency\_Savings_{ijt} = \beta_1 + \beta_2 AMI\_Adoption_{ijt} + \beta_3 C_{3,ijt} + \beta_4 C_{4,jt} + \gamma_t + \delta_{ij} + \varepsilon_{ijt}$

(2)

where $AMI\_Adoption_{ijt}$ represents the smart meter adoption rate for utility $i$ in state $j$ in year $t$ and $ARRA\_funding_{ijt}$ is a binary variable that equals one if the utility is a recipient of ARRA funding in year $t$, while $State\_Action_{jt}$ denotes the policy count related to advanced metering in state $j$ in the previous year $t-1$. $C_{1,ijt}$ represents the utility-level characteristics selected using the LASSO method, such as a wholesale dummy variable that takes the value of one if the utility operates in the wholesale market. $C_{2,jt}$ encompasses a set of state-level factors, such as income per capita, energy efficiency score, and electricity price. The term, $\mu_{ij}$, is the unobserved random effect that varies across utility-state combinations but not over time. Lastly, $\epsilon_{ijt}$ is an idiosyncratic error term. In equation (2), $Energy\_Efficiency\_Savings_{ijt}$ represents incremental annual energy savings from energy efficiency programs in the log term. $C_{3,ijt}$ is another set of utility-level controls including the log of the number of residential customers, the wholesale dummy variable, and the investor-owned dummy variable that equals one if the utility is investor-owned. $C_{4,jt}$ is a set of state-level factors, including income per capita, and energy intensity. $\gamma_t$ captures the year fixed effect, accounting for common changes across all utilities in year t that could influence their willingness to improve the energy efficiency savings. $\delta_{ij}$ represents the fixed effects specific to each utility-state combination, controlling for average differences between utilities and states. $\varepsilon_{ijt}$ is the disturbance term. The subscripts and other variables are the same as those in Equation (1). These model specifications aim to account for the complexities and potential sources of bias in the relationship between smart meter adoption, government policies, and energy efficiency savings. The inclusion of fixed effects and instrumental variables helps mitigate endogeneity concerns, while the selection of control variables using the LASSO method aids in addressing model complexity and overfitting issues.

## 4 Data description

This study draws on data from various sources spanning the period from 2007 to 2020. Table 1 displays the list of variables, basic statistics, their descriptions, and the respective data sources. The primary dataset used in this study is derived from the survey data collected by the EIA (EIA, 2007-2020). The EIA has been gathering information on the number of AMI (smart meters) and AMR as part of its Annual Electric Power Industry Report, Form EIA-861 since 2007 (EIA, 2007-2020). To measure the utility's energy efficiency savings, we utilize incremental annual energy savings, which capture the changes in aggregate electricity usage resulting from participation in energy efficiency programs. These savings are collected on an annual basis and are measured in Megawatt-hour. Since the data for this dependent variable exhibits a right-skewed distribution, we apply a natural logarithm transformation following the Box-Cox transformation techniques (Box & Cox, 1964) to reduce skewness. This transformation yields a nearly symmetrical distribution of the variable, facilitating the interpretation of the coefficients. After a thorough data cleaning process, we obtained a final analysis dataset comprising 20,038 observations from 2,798 electric utilities.

**Table 1.** Data description

| Variable | Mean | S.D. | Description | Data Source |
|---|---|---|---|---|
| Energy Efficiency Savings | 1.43 | 3.02 | Incremental Annual Energy Savings of Energy Efficiency Programs in Residential Sector (MWh) in log term | U.S. EIA |
| AMI Adoption | 32.75 | 45.44 | AMI meter adoption rate (%) in the residential sector | U.S. EIA |
| ARRA funding | 0.017 | 0.13 | Binary, =1 if received federal ARRA funding, = 0 otherwise | SmartGrid.gov |
| State Actions | 0.16 | 0.42 | State policy count (lagged one year) | Various Sources |
| Investor-Owned | 0.07 | 0.26 | Binary, =1 if investor-owned utility, = 0 otherwise | U.S. EIA |
| Wholesale | 0.11 | 0.31 | Binary, =1 if operates in the wholesale market | U.S. EIA |
| Residential Customer | 9.19 | 1.84 | The number of residential customers in log term | U.S. EIA |
| Income | 10.72 | 0.16 | Income per capita in $1000 | BEA |
| Energy Efficiency Score | 0.35 | 0.19 | State Energy Efficiency Scores (range from 0 to 1) | ACEEE |
| Energy Intensity | 7.16 | 2.39 | State total energy consumed per dollar of real GDP (Btu/$) | U.S. EIA |
| Electricity Price | 34.43 | 6.67 | Average State Electricity Price($ per million Btu) | U.S. EIA |
| N | 20,038 | | | |
| Year | 2007-2020 | | | |

We also use the total number of residential customers as a proxy for measuring the overall count of residential meters, which helps us calculate the proportion of AMI adoption. The adoption rate is represented as a continuous and non-negative variable, measured on a scale of 0 to 100. As illustrated in Table 1, the average AMI adoption rate stands at 32.75 percent, accompanied by a standard derivation of 45 percent. To gather data on ARRA funding, we referred to Smartgird.gov and energy.gov, both maintained by the DOE (DOE, 2021). These websites provide information on ARRA funding recipients, award amount, total project value, as well as the start and end years of ARRA funding. However, the specific annual funding amounts received by utilities remain unknown. Consequently, we have represented ARRA funding through a binary variable. This variable takes a value of one if a utility received ARRA funding during the 2010-2015 period, and zero otherwise. For instance, Central Maine Power Company received $95.9 million from ARRA between 2010 and 2014, resulting in a value of one for the ARRA funding variable during these years, and zero for the remaining years. This encoding method enables us to emulate an experimental research design, allowing us to examine the differential effect of the ARRA funding on utilities that received funding compared to those that did not. State policy data from 2007 to 2012 were extracted from S. Zhou and Matisoff (2016), while data from 2013 to 2020 were sourced from AMI reports and the Federal Energy Regulatory Commission's reports on demand response and advanced metering (DOE, 2020; FERC, 2019).[1]

Additionally, utility characteristics were incorporated into our analysis by using data from the EIA (EIA, 2007-2020). The frequency distributions of utility ownership types are presented in Table 2, providing an overview of the sample consisting of 2,798 utilities. Among these utilities, less than 5% are categorized as investor-owned, 28% are classified as co-operatives, and the majority, comprising 62%, are municipally owned.

Table 2. Frequency table for ownership type

| Ownership Type | Frequency | Percent | Cumulative Percent |
| --- | --- | --- | --- |
| Municipal-Owned | 1,754 | 62.69 | 62.69 |
| Co-operative | 800 | 28.59 | 91.28 |
| Investor-Owned | 131 | 4.68 | 95.96 |
| Other | 113 | 4.04 | 100 |

In our analysis, we incorporate controls to account for various factors. Firstly, we include the total number of residential customers as a control variable, as it serves as a measure of utility size. This is important because larger utilities have the potential to achieve economies of scale through the adoption of new technologies (Almus & Nerlinger, 1999). Moreover, we introduce a dummy variable that takes a value of one if the utility operates in the wholesale market, allowing us to account for any differences associated with the wholesale market operations. To further enhance our analysis, we employ additional data sources, such as ACEEE (ACEEE, 2007), Bureau of Economic Analysis (BEA, 2020), and NOAA (NOAA, 2020) to derive state-level control variables. These variables include the energy efficiency index, electricity price, energy intensity, and income per capita. By incorporating these variables, we can consider and control for various state-level factors that may influence AMI adoption and its associated outcomes.

---

[1] Please contact authors for a summary of state legislative actions concerning AMI.

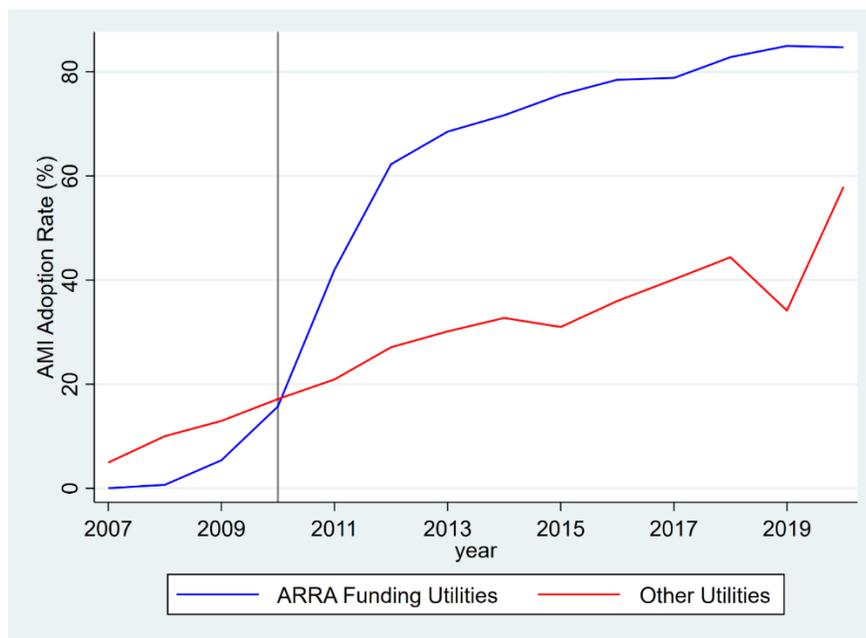

**Figure 3.** AMI Adoption Rate for Utilities with or without ARRA Funding, 2007-2020

To examine the potential differences in AMI adoption rate patterns before and after 2010 between utilities that received ARRA funding and those that did not, we divided our sample into two subsamples: the ARRA funding utility subsample and the non-ARRA funding utility subsample. Figure 4 illustrates the respective patterns of AMI adoption rate for each subsample. Upon analyzing the post-2010 period, we observed that the average adoption rate of ARRA funding utilities was much higher compared to non-ARRA funding utilities. Conversely, during the pre-2010 period when ARRA funding was granted, the pattern was the opposite. To further investigate the differences, we conducted a two-sample t-test to compare the mean adoption rates between utilities that received ARRA funding and those that did not. The results of the t-test indicate statistically significant differences in mean adoption rates between the two groups after 2010.

| Variables | (1) | (2) | (3) | (4) | (5) | (6) | (7) | (8) | (9) | (10) | (11) |
|---|---|---|---|---|---|---|---|---|---|---|---|
| (1) Energy_Efficiency_Savings | 1.000 | | | | | | | | | | |
| (2) AMI_Adoption | 0.061 | 1.000 | | | | | | | | | |
| (3) ARRA_Funding | 0.105 | 0.074 | 1.000 | | | | | | | | |
| (4) State Actions | 0.040 | -0.015 | 0.034 | 1.000 | | | | | | | |
| (5) Investor_Owned | 0.445 | -0.035 | 0.100 | 0.014 | 1.000 | | | | | | |
| (6) Wholesale | 0.418 | -0.059 | 0.096 | 0.063 | 0.463 | 1.000 | | | | | |
| (7) Residential Customer | 0.453 | 0.181 | 0.107 | 0.035 | 0.494 | 0.353 | 1.000 | | | | |
| (8) Income | 0.194 | 0.096 | -0.048 | 0.086 | 0.039 | 0.079 | -0.098 | 1.000 | | | |
| (9) Energy Efficiency Score | 0.260 | -0.077 | 0.037 | 0.170 | 0.084 | 0.185 | 0.026 | 0.408 | 1.000 | | |
| (10) Energy Intensity | -0.199 | 0.022 | -0.000 | -0.143 | -0.072 | -0.150 | -0.065 | -0.419 | -0.654 | 1.000 | |
| (11) Electricity Price | 0.183 | 0.007 | 0.015 | 0.123 | 0.146 | 0.125 | 0.011 | 0.563 | 0.552 | -0.525 | 1.000 |

**Table 3.** Correlation Matrix

Table 3 presents the correlation matrix for the variables under consideration. Energy intensity exhibits a negative linear relationship with both the energy efficiency score and income per capita. On the other hand, the energy efficiency score shows a positive correlation with electricity prices. In our analysis, we implement Lasso regression to select the most relevant covariates for the first- and second-stage models. For the first stage model, the selected covariates are the wholesale dummy variable, income per capita, energy efficiency score, and electricity price. In the second stage models, the chosen covariates are the wholesale dummy variable, the number of residential customers, income per capita, and energy intensity.

## 5 Results

The estimated results from Equations (1) and (2) are displayed in Table 4. Model 1 presents the marginal effects in random effects Tobit regression. Focusing on the determinants of AMI adoption in the first-stage model, the coefficient for ARRA funding is 4.030 and statistically significant. This indicates that, when comparing utilities that received ARRA funding, the AMI adoption rates for the former are at least 4.03 percent higher, while holding all other factors constant. These findings provide support for Hypothesis 1, suggesting that ARRA funding has played a role in promoting the penetration of smart meters.

The results also indicate that state actions have an impact of 0.449 and are statistically significant. More importantly, the interaction term between ARRA funding and State Actions is positive and statistically significant at the 0.05 level. These findings highlight important relationships. Specifically, for utilities that did not receive ARRA funding, each additional state action is associated with an expected increase in the AMI adoption rate of 0.449%. However, for utilities that received ARRA funding, each additional state legislative action is linked to a substantial increase in the expected AMI adoption of approximately 2.44%. These results suggest that the adoption rate is generally higher for utilities that received ARRA funding. Furthermore, the adoption rate tends to increase at a higher rate with an increase in the number of state actions for utility companies that received federal funding compared to those that did not. Overall, these findings support Hypotheses 2 and 3, which posit a positive relationship between the adoption rate and state actions, as well as between the adoption rate and the interactions of state actions and ARRA funding.

The estimated regression results, as displayed in Model 2 of Table 4, provide insights into the relationship between the dependent variable of energy efficiency savings and various factors. After addressing the issue of endogeneity, the results reveal a positive and statistically significant impact of the smart meter adoption rate on annual energy efficiency savings. The estimated coefficient of 0.012 is statistically significant at the 0.01 level. Specifically, the findings suggest that a one percent increase in AMI adoption leads to a 1.2 percent increase in incremental energy efficiency savings, measured in megawatt-hours. These empirical results provide support for our fourth hypothesis, indicating that an increase in smart meter adoption is associated with an increase in utilities' energy efficiency savings.

Among the other control variables included in the second stage estimation, the coefficient for the investor-owned dummy variable is positive, indicating a potential association with higher savings in energy efficiency programs compared to non-investor-owned utilities. However, it is important to note that this association is statistically insignificant. And the wholesale dummy variable exhibits a positive and significant coefficient. This suggests that wholesalers benefit from economies of scale when it comes to energy savings. Additionally, the coefficients for state-level factors, such as income per capita, and energy intensity, are positive and significant. This confirms our expectations that higher income levels and increased energy consumption contribute to improved utility energy efficiency program performance. Table 4 demonstrates that the F-tests indicate statistical significance for the second-stage model. Furthermore, the regressors in the second stage contribute to approximately 23% of the variation in the annual energy efficiency savings. These results underscore the significance of the included variables in explaining the variation in both smart meter adoption and energy efficiency savings.

To ensure the robustness of our results, we conducted additional robustness tests using various modeling approaches. These included random effects models, fixed effects models, and first-differences models.[2] The result of these robustness tests consistently supports and confirms the findings obtained from our initial models. This indicates that the observed relationships and effects are robust even after accounting for the variation across utilities and considering a one-period change for each company. We have also strengthened the validity and reliability of our results, further supporting the robustness of our findings concerning the factors influencing smart meter adoption and energy efficiency savings.

---

[2] The results are not present in the paper but are available upon request.

| Variables | Model 1: First Stage Random Effects Tobit Regression AMI Adoption | Model 2: Second Stage Fixed Effects Regression Total Energy Efficiency Savings |
|---|---|---|
| ARRA Funding | **4.030**\*\*\* | |
| | (0.601) | |
| State Actions | **0.449**\*\* | |
| | (0.211) | |
| ARRA Funding * State Action | **1.991**\*\* | |
| | (0.874) | |
| AMI Adoption | | **0.012**\*\*\* |
| | | (0.003) |
| Investor-Owned | | 0.126 |
| | | (1.36) |
| Wholesale | 1.167 | 0.276* |
| | (0.965) | (0.160) |
| Residential Customer | | 0.557*** |
| | | (0.174) |
| Income | 48.746*** | 1.001 |
| | (1.388) | (1.061) |
| Energy Efficiency Score | -0.056* | |
| | (0.029) | |
| Electricity Price | 0.179*** | |
| | (0.042) | |
| Energy Intensity | | 0.233*** |
| | | (0.051) |
| Constant | | -16.39 |
| | | (11.69) |
| Year FE | | YES |
| Utility by State FE | | YES |
| Random Effects | Yes | |
| Observations | 20,038 | 20,038 |
| R-squared | --- | 0.23 |
| Model F | --- | 218.82*** |

Table 4 Regression Results

Standard errors are reported in parentheses, *** $p<0.01$, ** $p<0.05$, *$p<0.1$

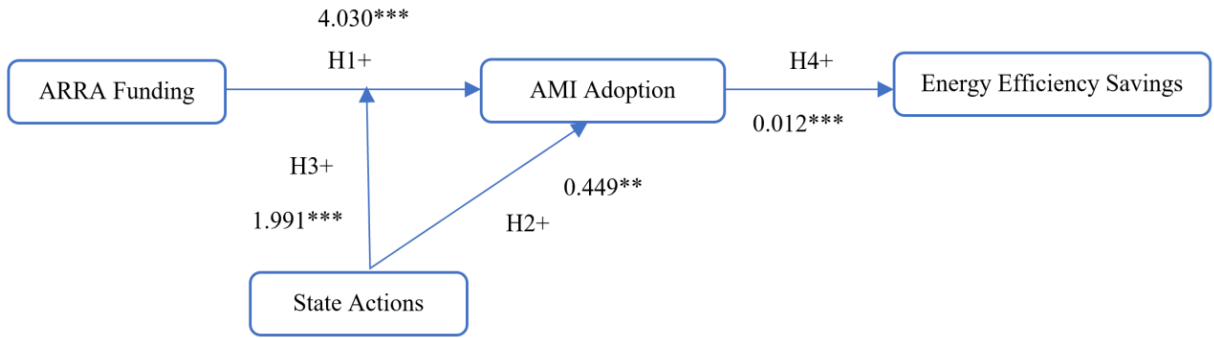

**Figure 4.** Summary of Results. (*** Significant at p=0.01 level)

# 6 Discussion and policy implications

In conclusion, as shown in Figure 5, our study provides robust empirical evidence for the significant impacts of federal ARRA funding and state policies on smart meter adoption. The combined impact of these two policy instruments is found to have a significantly greater impact on smart meter adoption than the sum of their individual interventions alone. This highlights the importance of considering the synergistic effects of multiple policy interventions. Furthermore, we find that a higher smart meter adoption rate is associated with greater energy efficiency savings, after accounting for endogeneity. The estimated beta coefficient of 0.044 suggests that a 1% increase in smart meter adoption rate is connected with a 4.8% increase in annual energy efficiency savings, which is equivalent to approximately 2,300 KWh or 2.3 MWh on average. These findings align with previous studies that highlight the positive impact of smart meter usage on energy savings. By enabling utilities and energy consumers to make more informed choices, smart meters contribute to reducing energy consumption without sacrificing services. Overall, our study underscores the importance of federal funding and state policies in promoting smart meter adoption and highlights the potential for energy efficiency gains associated with these advancements in technology.

This paper makes contributions to environmental policy research in two key areas. Firstly, it highlights the importance of adopting a multilevel governance perspective in future studies. Existing research often neglects the interactive effects of policies at multiple levels of government and the few studies that have explored this perspective lacked up-to-date data at the utility level. By incorporating federal funding and state policies into the analysis, this study demonstrates the synergistic effects of these interventions in promoting smart meter adoption. This contributes to the development of a more comprehensive understanding of the multilevel governance approach in environmental policy research. Secondly, this study is one of the pioneering works to provide empirical evidence for the positive association between smart meter adoption and energy efficiency savings in the residential sector. While previous research has predominantly focused on examining the factors influencing smart meter adoption, limited attention has been given to the actual energy efficiency outcomes. By establishing a positive relationship between smart meter adoption and energy efficiency savings, this study extends the existing literature and provides empirical support for the potential energy-saving benefits of smart meter technologies. The robustness of the findings further strengthens the contribution of this study. The consistent sign, magnitude, and significance of the coefficients across different estimation methods validate the reliability of the results.

Our findings have important policy implications, providing a rationale for policy interventions aimed at promoting smart meter adoption and offering insights for further policy development and assessment. Hypothesis 4 suggests a positive relationship between smart meter adoption and energy savings. This improvement in energy efficiency can be attributed to the enhanced sensing and information processing capacities provided by smart devices. With access to detailed consumption data and patterns through smart meters, utility companies may have been able to make more informed decisions to improve their internal operations and optimize energy usage. Additionally, the installation of smart meters may have empowered consumers to actively participate in energy efficiency programs offered by utility companies. This engagement could result in energy savings through the replacement of outdated electric appliances and HVAC systems with high-efficiency alternatives. While our findings suggest a positive association between smart meter adoption and energy efficiency savings, it is important to note that further analysis is necessary to identify the underlying mechanisms through which these benefits are generated. Additionally establishing a causal relationship between smart meter adoption and energy efficiency savings would require more comprehensive research and rigorous study designs.

Our study provides insights into the effectiveness of government interventions, particularly through a multilevel governance approach. While a nationwide compulsory deployment plan for smart meters is politically challenging in the U.S. (S. Zhou & Matisoff, 2016), our findings suggest that government interventions in the form of economic stimulus or financial assistance programs can play a crucial role in achieving important goals in the absence of a federal mandate. Since the completion of the ARRA program, there have been limited visible federal actions in promoting smart meter adoption However, in light of recent developments, such as the Biden administration's announcement of a $2 trillion plan to build a modern and sustainable infrastructure, there is a potential for continued investment from the federal government. For instance, the DOE launched the 'Building a Better Grid' initiative in January 2022, which includes a deployment of $20 billion in federal financial tools, including $2 billion in extensions of the Smart Grid Investment Grant Program. Such public investment in smart grid technologies has the potential to yield long-term benefits across multiple dimensions.

Additionally, this study highlights the significant role of state legislative activities in promoting smart meter adoption and emphasizes the synergistic relationship between state legislation and federal funding. The findings provide statistical evidence to support the involvement of state legislatures and PUCs in driving smart meter adoption. State regulators can play a crucial role in promoting smart meter adoption by implementing various measures. For example, they can require utilities to establish comprehensive cost-recovery mechanisms for smart meter investments, ensuring that the full capabilities of these technologies are leveraged. In some cases, regulators may demand utilities revise and improve their plans to maximize customer engagement and ensure the effective implementation of smart meters. State regulators can also create performance incentives as part of the smart meter approval process, encouraging utilities to demonstrate how advanced meters can be utilized to achieve energy efficiency and savings. Furthermore, statewide marketing, customer education, and outreach programs can be implemented to inform residential customers about the benefits of smart meters and how they can optimize their energy consumption. These state-level actions, when combined with federal funding, can create a supportive environment for smart meter adoption and foster greater energy efficiency in the residential sector.

While our research focuses on the U.S. context, the findings and insights obtained can have broader implications for multi-level governance in other countries or regions, particularly in the development and deployment of smart meter technologies and climate change policymaking. The European Union (EU) serves as an example of a region where multi-level governance plays an important role in addressing climate change challenges. The EU's governance structure combines intergovernmental collaboration between member states with supranational integration. This multi-level approach ensures that actions taken at the EU level are justified based on the participation and coordination of national, regional, and local levels.

It is important to acknowledge the limitations of our study. First, the observational nature of the study introduces the possibility of biases and confounding effects, which may limit the ability to establish a causal relationship between multilevel governance, smart meter adoption, and utilities' energy performance. While we have taken measures to address endogeneity and conducted robustness tests, there may still be unobserved factors that could influence the results. Therefore, caution is needed in drawing definitive conclusions regarding causality. To gain a deep understanding of the mechanisms by which energy efficiency is enhanced through smart meter adoption and how governmental policies drive this adoption, a more qualitative and experimental approach could be beneficial. This would allow for a more in-depth exploration of the interactions and dynamics between various stakeholders involved in the smart meter deployment process. Additionally, incorporating the interaction between industry and government efforts would provide further insights. If data on industry-government collaborations are available, analyzing the dynamic interactions between the public and private partners could offer a more comprehensive understanding of the factors influencing smart meter adoption and energy performance outcomes.

# 7 Conclusions

In conclusion, our study provides empirical evidence on the relationships between multi-level governance, residential smart meter adoption, and energy efficiency performance. The results highlight the positive associations between federal funding, state legislative actions, and smart meter adoption in the residential sector. Moreover, the combined effect of these policy interventions is greater than the individual effects. The adoption, in turn, positively contributes to energy efficiency savings. We find that government economic incentives, coupled with state regulatory activities, lead to more desirable results for smart meter adoption, and thereafter, for improvements in energy performance in the utility industry.

**Declaration of generative AI and AI-assisted technologies in the writing process**

During the preparation of this work the authors used ChatGPT in order to improve readability and language. After using this tool/service, the authors reviewed and edited the content as needed and take full responsibility for the content of the publication.

**References**

[dataset]ACEEE (American Council for an Energy-Efficient Economy) (2021). *The State Energy Efficiency Scorecard for 2007-2020*. Retrieved from https://www.aceee.org/state-policy/scorecard/. Accessed June 19, 2021

ACEEE (American Council for an Energy-Efficient Economy) (2020). *Leveraging Advanced Metering Infrastructure To Save Energy*. Retrieved from https://www.aceee.org/sites/default/files/publications/researchreports/u2001.pdf/. Accessed August 1, 2021

Almus, M., & Nerlinger, E. A. (1999). Growth of new technology-based firms: which factors matter? *Small business economics, 13*(2), 141-154.

An, E. A. U. (2011). Smart Meters and Smart Meter Systems: A Metering Industry Perspective.

Andersson, K. P., & Ostrom, E. (2008). Analyzing decentralized resource regimes from a polycentric perspective. *Policy sciences, 41*(1), 71-93.

Arif, A., Al-Hussain, M., Al-Mutairi, N., Al-Ammar, E., Khan, Y., & Malik, N. (2013). *Experimental study and design of smart energy meter for the smart grid.* Paper presented at the 2013 International Renewable and Sustainable Energy Conference (IRSEC).

Bakker, K., & Ritts, M. (2018). Smart Earth: A meta-review and implications for environmental governance. *Global environmental change, 52*, 201-211.

[dataset] BEA (U.S. Bureau of Economic Analysis) (2020). U. S. *Personal Income Summary: Personal Income, Population, Per Capita Personal Income*. Retrieved from: https://apps.bea.gov/iTable/iTable.cfm?reqid=70&step=1&acrdn=2/. Accessed May 1, 2021

Belloni, A., Chernozhukov, V., & Hansen, C. (2014). High-dimensional methods and inference on structural and treatment effects. *Journal of Economic Perspectives, 28*(2), 29-50.

Box, G. E., & Cox, D. R. (1964). An analysis of transformations. *Journal of the Royal Statistical Society: Series B (Methodological), 26*(2), 211-243.

California Legislative Counsel. (2009-2010). *Senate Bill 1476 of the 2009-2010 Session*. Retrieved from http://www.leginfo.ca.gov/pub/09-10/bill/sen/sb_1451-1500/sb_1476_bill_20100929_chaptered.pdf

Corbett, J., Wardle, K., & Chen, C. (2018). Toward a sustainable modern electricity grid: The effects of smart metering and program investments on demand-side management performance in the US electricity sector 2009-2012. *IEEE Transactions on Engineering Management, 65*(2), 252-263.

del Río, P., & Cerdá, E. (2017). The missing link: The influence of instruments and design features on the interactions between climate and renewable electricity policies. *Energy Research & Social Science, 33*, 49-58.

Depuru, S. S. S. R., Wang, L., Devabhaktuni, V., & Gudi, N. (2011). *Smart meters for power grid—Challenges, issues, advantages and status*. Paper presented at the 2011 IEEE/PES Power Systems Conference and Exposition.

Di Gregorio, M., Fatorelli, L., Paavola, J., Locatelli, B., Pramova, E., Nurrochmat, D. R., ... & Kusumadewi, S. D. (2019). Multi-level governance and power in climate change policy networks. *Global environmental change*, *54*, 64-77.

DOE (U.S. Department of Energy) (2010). Communications requirements of Smart Grid technologies. *US Department of Energy, Tech. Rep*, 1-69.

DOE (U.S. Department of Energy) (2016). *Advanced Metering Infrastructure and Customer Systems: Results from the Smart Grid Investment Grant Program*. Retrieved from https://www.energy.gov/sites/prod/files/2016/12/f34/AMI%20Summary%20Report_09-26-16.pdf/. Accessed June 19, 2020

DOE (U.S. Department of Energy) (2020). *AMI IN REVIEW*. Retrieved from smartgrid.gov: https://www.smartgrid.gov/files/documents/AMI_Report_7_8_20_final_compressed.pdf/. Accessed June 19, 2021

[dataset] DOE (U.S. Department of Energy) (2021). The Smart Grid. Retrieved from https://www.smartgrid.gov/the_smart_grid/smart_grid.html/. Accessed May 21, 2021

Drukker, D., & Liu, D. (2019). An Introduction to the Lasso in Stata. *The Stata Blog*.

[dataset] EIA (U.S. Energy Information Administration) (2007-2020). *Annual Electric Power Industry Report, Form EIA-861 detailed data files*. Retrieved from: https://www.eia.gov/electricity/data/eia861/. Accessed June 20, 2021

EIA (U.S. Energy Information Administration) (2011). Smart Grid Legislative and Regulatory Policies and Case Studies. Retrieved from: https://www.eia.gov/analysis/studies/electricity/. Accessed May 25, 2021

EIA (U.S. Energy Information Administration) (2017). Nearly half of all U.S. electricity customers have smart meters. Retrieved from https://www.eia.gov/todayinenergy/detail.php?id=34012/. Accessed December 10, 2021

EIA (U.S. Energy Information Administration) (2020). How many smart meters are installed in the United States, and who has them? Retrieved from https://www.eia.gov/tools/faqs/faq.php?id=108&t=1/. Accessed December 15, 2021

EIA (U.S. Energy Information Administration) (2020). *Electric Sales, Revenue, and Average Price*. Retrieved from: https://www.eia.gov/electricity/sales_revenue_price/

EPA (U.S. Environmental Protection Agency) (2014). Electricity Customers. Retrieved from https://www.epa.gov/energy/electricity-customers

EPR (Energy Policy Review) (2008). *Energy Policies of IEA Countries: United States 2007*. Retrieved from Energy Policy Review: https://www.iea.org/reports/energy-policies-of-iea-countries-united-states-2007/. Accessed December 20, 2021

EPRI (Electric Power Research Institute) (2007). *The Power to Reduce CO2 Emissions*. Retrieved from: http://mydocs.epri.com/docs/public/DiscussionPaper2007.pdf

ESGTF (The European Smart Grid Task Force) (2010). *Communication From the Commission to the European Parliament, the Council, the European Economic and Social Committee and the Committee of the Regions.* Retrieved from https://op.europa.eu/en/publication-detail/-/publication/ac9cd214-53c6-11ea-aece-01aa75ed71a1/language-en/. Accessed December 15, 2021

Faruqui, A., Hledik, R., & Sergici, S. (2009). Piloting the smart grid. *The Electricity Journal, 22*(7), 55-69.

Feldman, D. (2016). Polycentric Governance.
Feng, S., Zhang, J., & Gao, Y. (2016). Investment uncertainty analysis for smart grid adoption: A real options approach. *Information Polity, 21*(3), 237-253.
FERC (U.S. Federal Energy Regulatory Commission) (2010). *2010 Assessment of Demand Response and Advanced Metering Report*. Retrieved from https://www.ferc.gov/sites/default/files/2020-04/2010-dr-report.pdf/. Accessed December 12, 2021
FERC (U.S. Federal Energy Regulatory Commission) (2019). *2019 Assessment of Demand Response and Advanced Metering Report*. Retrieved from https://www.ferc.gov/industries-data/electric/power-sales-and-markets/demand-response/reports-demand-response-and/. Accessed December 12, 2021
FERC (U.S. Federal Energy Regulatory Commission) (2022). What FERC Does. Retrieved from https://www.ferc.gov/about/what-ferc/what-ferc-does/. Accessed January 12, 2022
GAO (U.S. Government Accountability Office) (2011). 'Electricity Grid Modernization: Progress Being Made on Cybersecurity Guidelines, but Key Challenges Remain to be Addressed'. In: US Government Accountability Office.
Gao, Y., Fang, C., & Zhang, J. (2022). A Spatial Analysis of Smart Meter Adoptions: Empirical Evidence from the U.S. Data. *Sustainability, 14*(3), 1126.
Gao, Y., & Zhang, J. (2021). *Studying the Impacts of Federal Funding on Residential Smart Meter Adoption and Utilities' Performance in the U.S.: A Simultaneous Equation Approach*. Paper presented at the DG.O2021: The 22nd Annual International Conference on Digital Government Research, Omaha, NE, USA. https://doi.org/10.1145/3463677.3463739
Gellings, C. (2011). Estimating the costs and benefits of the smart grid: a preliminary estimate of the investment requirements and the resultant benefits of a fully functioning smart grid. *Electric Power Research Institute (EPRI), Technical Report (1022519), 1*.
Giest, S. (2020). Do nudgers need budging? A comparative analysis of European smart meter implementation. *Government Information Quarterly, 37*(4), 101498.
Gil-Garcia, J. R., Zhang, J., & Puron-Cid, G. (2016). Conceptualizing smartness in government: An integrative and multi-dimensional view. *Government Information Quarterly, 33*(3), 524-534.
Goulder, L. H., & Stavins, R. N. (2011). Challenges from state-federal interactions in US climate change policy. *American Economic Review, 101*(3), 253-257.
Gungor, V. C., Sahin, D., Kocak, T., Ergut, S., Buccella, C., Cecati, C., & Hancke, G. P. (2011). Smart grid technologies: Communication technologies and standards. *IEEE transactions on Industrial informatics, 7*(4), 529-539.
Guo, C., Bond, C. A., & Narayanan, A. (2015). *The adoption of new smart-grid technologies: incentives, outcomes, and opportunities*: Rand Corporation.
Guo, P., Li, V. O., & Lam, J. C. (2017). Smart demand response in China: Challenges and drivers. *Energy Policy, 107*, 1-10.
Horbach, J., Rammer, C., & Rennings, K. (2012). Determinants of eco-innovations by type of environmental impact—The role of regulatory push/pull, technology push and market pull. *Ecological economics, 78*, 112-122.
Howlett, M., Vince, J., & Río González, P. d. (2017). Policy integration and multi-level governance: dealing with the vertical dimension of policy mix designs.

Jaffe, A. B., Newell, R. G., & Stavins, R. N. (2005). A tale of two market failures: Technology and environmental policy. *Ecological economics, 54*(2-3), 164-174.
Jaffe, A. B., & Stavins, R. N. (1994). The energy paradox and the diffusion of conservation technology. *Resource and Energy Economics, 16*(2), 91-122.
James, G., Witten, D., Hastie, T., & Tibshirani, R. (2013). *An introduction to statistical learning* (Vol. 112): Springer.
Kennedy, P. (2008). *A guide to econometrics*: John Wiley & Sons.
Lang, C., & Okwelum, E. (2015). The mitigating effect of strategic behavior on the net benefits of a direct load control program. *Energy Economics, 49*, 141-148.
Mah, D. N.-y., van der Vleuten, J. M., Ip, J. C.-m., & Hills, P. R. (2012). Governing the transition of socio-technical systems: a case study of the development of smart grids in Korea. *Energy Policy, 45*, 133-141.
Marks, G. (1993). Structural policy and multilevel governance in the EC. *The Maastricht debates and beyond, 392*.
MassSave. (2022). Heat Pump Rebates. Retrieved from https://www.masssave.com/en/saving/residential-rebates/heat-pump/. Accessed December 12, 2021
Matisoff, D. C. (2008). The adoption of state climate change policies and renewable portfolio standards: Regional diffusion or internal determinants? *Review of Policy Research, 25*(6), 527-546.
Moura, P. S., López, G. L., Moreno, J. I., & De Almeida, A. T. (2013). The role of Smart Grids to foster energy efficiency. *Energy Efficiency, 6*(4), 621-639.
NIST (U.S. National Institute of Standards and Technology) (2021). Smart Grid Interoperability Panel. Retrieved from https://www.nist.gov/programs-projects/smart-grid-national-coordination/smart-grid-interoperability-panel-sgip/. Accessed December 18, 2021
NOAA (U.S. National Oceanic and Atmospheric Administration) (2020). *Degree Days Statistics*. Retrieved from: ftp://ftp.cpc.ncep.noaa.gov/htdocs/degree_days/weighted/legacy_files/. Accessed September 9, 2021
Norberg-Bohm, V. (2000). Creating incentives for environmentally enhancing technological change: lessons from 30 years of US energy technology policy. *Technological forecasting and social change, 65*(2), 125-148.
Ockwell, D. G., Watson, J., MacKerron, G., Pal, P., & Yamin, F. (2008). Key policy considerations for facilitating low carbon technology transfer to developing countries. *Energy Policy, 36*(11), 4104-4115.
Olmos, L., Ruester, S., Liong, S.-J., & Glachant, J.-M. (2011). Energy efficiency actions related to the rollout of smart meters for small consumers, application to the Austrian system. *Energy, 36*(7), 4396-4409.
Ostrom, E. (2009). A polycentric approach for coping with climate change. *Available at SSRN 1934353*.
Ostrom, V., Tiebout, C. M., & Warren, R. (1961). The organization of government in metropolitan areas: a theoretical inquiry. *American political science review, 55*(4), 831-842.
Podgornik, A., Sucic, B., Bevk, P., & Stanicic, D. (2013). The impact of smart metering on energy efficiency in low-income housing in Mediterranean. In *Climate-Smart Technologies* (pp. 597-614): Springer.

Power Technology, (2020). EU states approve $1.17bn investment in energy infrastructure. Retrieved from https://www.power-technology.com/news/european-union-eu-approves-energy-infrastructure-investment-connecting-europe-facility/. Accessed December 16, 2021

Samuels, J., & Booysen, M. J. (2019). Chalk, talk, and energy efficiency: Saving electricity at South African schools through staff training and smart meter data visualisation. *Energy Research & Social Science, 56*, 101212.

SCG (Structure Consulting Group, LLC.). (2010). *PG&E Advanced Metering Assessment Report* Retrieved from https://www.pge.com/includes/docs/pdfs/myhome/customerservice/meter/smartmeter/StructureReport.pdf/. Accessed December 12, 2021

Schaffrin, A., Sewerin, S., & Seubert, S. (2015). Toward a comparative measure of climate policy output. *Policy Studies Journal, 43*(2), 257-282.

Shea, D., & Bell, K. (2019). Smart Meter Opt-Out Policies. Retrieved from https://www.ncsl.org/research/energy/smart-meter-opt-out-policies.aspx/. Accessed December 5, 2021

Shobe, W. M., & Burtraw, D. (2012). Rethinking environmental federalism in a warming world. *Climate Change Economics, 3*(04), 1250018.

Sovacool, B. K., Hook, A., Sareen, S., & Geels, F. W. (2021). Global sustainability, innovation and governance dynamics of national smart electricity meter transitions. *Global Environmental Change*, *68*, 102272.

New York Public Service Commission. (2010). *In the Matter of the American Recovery and Reinvestment Act of 2009 - Utility Filings for New York Economic Stimulus.* State of New York Public Service Commission Retrieved from file:///C:/Users/Cryst/Downloads/%7B42972870-FE78-4CF3-BEB5-AB76D35DDF07%7D%20(1).pdf

Strong, D. R. (2019). Impacts of diffusion policy: determinants of early smart meter diffusion in the US electric power industry. *Industrial and Corporate Change, 28*(5), 1343-1363.

The Commonwealth Fund (2020). Federalism, the Affordable Care Act, and Health Reform in the 2020 Election. Retrieved from https://www.commonwealthfund.org/publications/fund-reports/2019/jul/federalism-affordable-care-act-health-reform-2020-election

U.S. Congress (2005). *Energy Policy Act of 2005.* Paper presented at the U.S. Congress.

U.S. Congress (2007). Energy independence and security act of 2007. *Public law, 2*, 110-140.

Viscusi, W. K., & Hamilton, J. T. (1999). Are risk regulators rational? Evidence from hazardous waste cleanup decisions. *American Economic Review, 89*(4), 1010-1027.

Globe New Wire. (2020). *China Smart Grid Network Market - Growth, Trends, and Forecasts (2020 - 2025)*. Retrieved from Reportlinker.com: https://www.reportlinker.com/p05989483/China-Smart-Grid-Network-Market-Growth-Trends-and-Forecasts.html?utm_source=GNW/. Accessed November 12, 2020

Zhang, D., Pee, L., Pan, S. L., & Cui, L. (2022). Big data analytics, resource orchestration, and digital sustainability: A case study of smart city development. *Government Information Quarterly, 39*(1), 101626.

Zhang, F., & Gallagher, K. S. (2016). Innovation and technology transfer through global value chains: Evidence from China's PV industry. *Energy Policy, 94*, 191-203.

Zhang, J., Luna-Reyes, L. F., Pardo, T. A., & Sayogo, D. S. (2016). *Information, Models, and Sustainability*: Springer.
Zhou, L., Xu, F.-Y., & Ma, Y.-N. (2010). *Impact of smart metering on energy efficiency.* Paper presented at the 2010 International Conference on Machine Learning and Cybernetics.
Zhou, S., & Brown, M. A. (2017). Smart meter deployment in Europe: A comparative case study on the impacts of national policy schemes. *Journal of Cleaner Production, 144*, 22-32.
Zhou, S., & Matisoff, D. C. (2016). Advanced metering infrastructure deployment in the United States: The impact of polycentric governance and contextual changes. *Review of Policy Research, 33*(6), 646-665.

**Table 5.** Literature Review Summary

| Variables | Measures | Reference |
|---|---|---|
| The Policy Determinants of Smart Meter Adoption | | |
| Federal Funding | Per capita ARRA funding allocated to AMI projects | Zhou & Matisoff (2016) Gao, et al (2022) |
| | Binary, =1 if utility received ARRA funding | Gao & Zhang (2021) |
| State policy or local policy | 1) Number of state AMI promotion policies 2) Number of state AMI data security and privacy policies | Zhou & Matisoff (2016) |
| | Binary, =1 if subject to state support for adoption of AMI | Strong (2019) |
| | Number of State AMI policies | Gao, et al (2022) |
| The Impacts of Smart Meter Adoption on Energy Efficiency | | |
| Energy efficiency | Reduction in customer electric bill | Zhou, Xu & Ma (2010) |
| | Reduction in electricity consumption | Olmos, et al (2011) |
| | Reduction in energy consumption | Arif, et al (2013) |
| | Electricity cost savings (%) | Podgornik, et al (2013) |
| | Energy efficiency savings (MWh) | Corbett, et al (2018) |
| | Electricity cost savings (%) | Samuels & Booysen (2019) |